\documentclass[pdflatex,sn-mathphys-num]{sn-jnl}
\usepackage{graphicx}%
\usepackage{multirow}%
\usepackage{amsmath,amssymb,amsfonts}%
\usepackage{amsthm}%
\usepackage{mathrsfs}%
\usepackage[title]{appendix}%
\usepackage{xcolor}%
\usepackage{textcomp}%
\usepackage{manyfoot}%
\usepackage{booktabs}%
\usepackage{algorithm}%
\usepackage{algorithmicx}%
\usepackage{algpseudocode}%
\usepackage{listings}%
\usepackage{tikz}%
\usepackage{t1enc}
\usepackage[utf8]{inputenc}

\begin{document}

\title[Lévy walk of pions in heavy-ion collisions]{Lévy walk of pions in heavy-ion collisions}

\author*[1]{\fnm{Dániel} \sur{Kincses}}\email{kincses@ttk.elte.hu}
\equalcont{These authors contributed equally to this work.}
\author[1]{\fnm{Márton} \sur{Nagy}}\email{marton.nagy@ttk.elte.hu}
\equalcont{These authors contributed equally to this work.}
\author[1]{\fnm{Máté} \sur{Csanád}}\email{csanad@elte.hu}    
\equalcont{These authors contributed equally to this work.}

\affil[1]{\orgdiv{Department of Atomic Physics}, \orgname{ELTE Eötvös Loránd University}, \orgaddress{\street{Pázmány Péter sétány 1/A}, \city{Budapest}, \postcode{H-1117}, \country{Hungary}}}

\abstract{The process of Lévy walk, i.e., movement patterns described by heavy-tailed random walks, plays a role in various phenomena, from chemical and microbiological systems through marine predators to climate change. Recent experiments have suggested that this phenomenon also appears in heavy-ion collisions. However, the theoretical interpretation supporting such findings is still debated. In high-energy collisions of heavy nuclei, the strongly interacting Quark Gluon Plasma is created, which, similarly to the early Universe, undergoes a rapid expansion and transition back to hadronic matter. In the subsequent expanding hadron gas, particles interact until kinetic freeze-out, when their momenta stop changing, and they freely transition toward the detectors. Measuring spatial freeze-out distributions is a crucial tool in understanding the dynamics of the created matter and the interactions among its constituents. In this paper, we introduce a three-dimensional analysis of the spatial freeze-out distribution of pions (the most abundant particles in such collisions). Utilising Monte-Carlo simulations of high-energy collisions, we show that the chain of processes ending in a final state pion has a step length distribution leading to Lévy-stable distributions. Subsequently, we show that simulated pion freeze-out distributions indeed exhibit heavy tails and can be described by a three-dimensional elliptically contoured symmetric Lévy-stable distribution.}

\keywords{Lévy walk, stable distribution, high-energy physics, heavy-ion physics, femtoscopy}

\maketitle

\section{Introduction}\label{sec1}

\begin{figure}
    \centering
    \includegraphics[width=\textwidth]{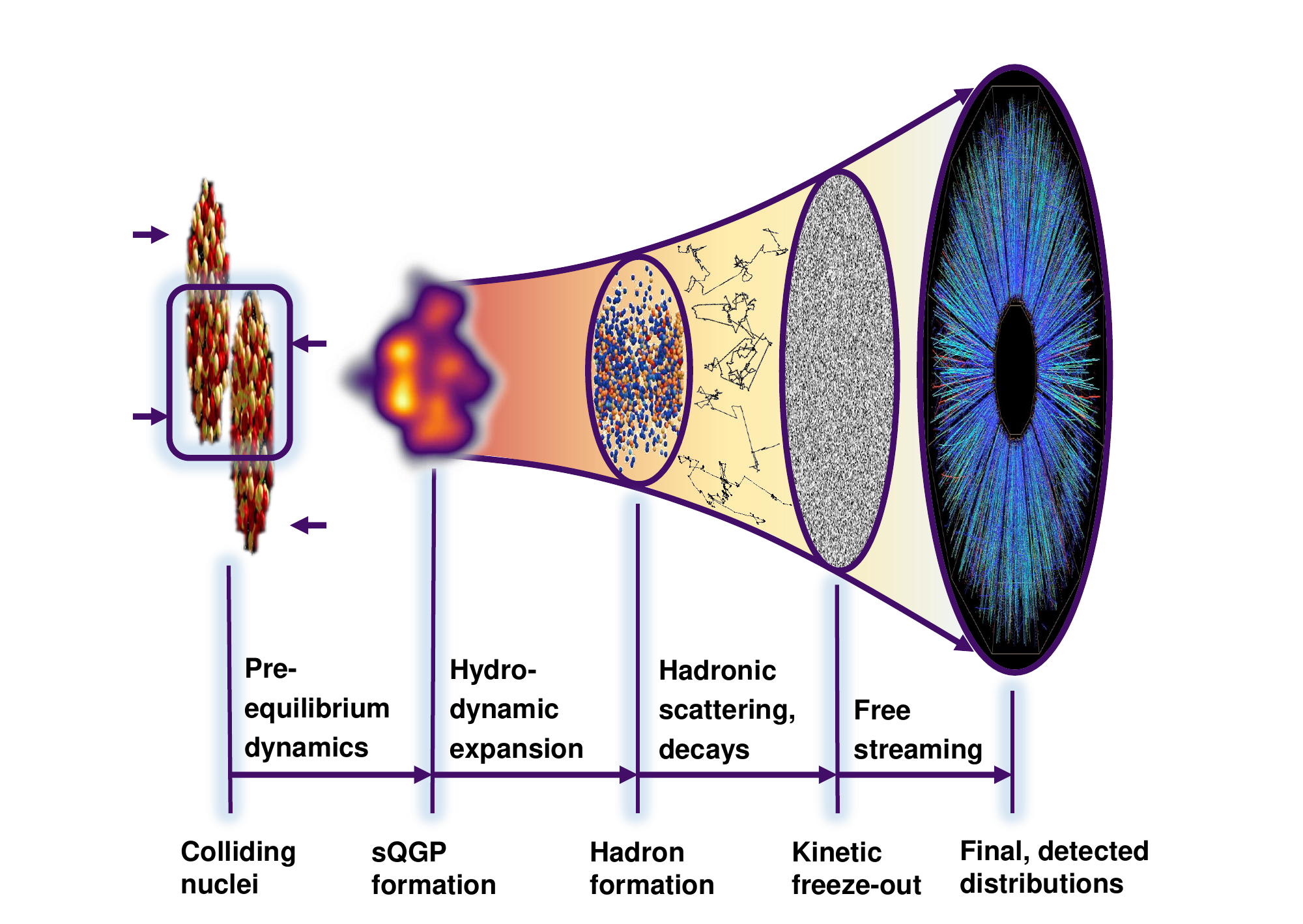}
    \caption{An illustration of a heavy-ion collision. 
    The stages of the time evolution are marked by horizontal arrows and separated by vertical lines, indicating changes in the characteristics of the system. The illustration of the initial-state fluctuations are taken from Ref.~\cite{Bernhard:2019bmu}. Figure adapted from Ref.~\cite{Sorensen:2009cz}.}
    \label{f:illustration}
\end{figure}

High-energy physics aims to understand the fundamental constituents of the Universe and their interactions. One branch of high-energy physics is heavy-ion physics, which seeks to understand the strong interaction and the medium governed by it, formed by quarks and gluons. One form of this medium is ordinary nuclear matter. Another one is the hadron gas, where various particles (baryons and mesons) form a loosely interacting gas. Yet another one can be created in ultra-relativistic heavy-ion collisions (illustrated in Fig.~\ref{f:illustration}): this is the so-called strongly interacting Quark-Gluon Plasma (sQGP), observed at the Super Proton Synchrotron (SPS), Relativistic Heavy Ion Collider (RHIC) and Large Hadron Collider (LHC) particle accelerators~\cite{Adcox:2004mh,Adams:2005dq,Back:2004je,Arsene:2004fa,Aamodt:2010jd,Aamodt:2010pa,CMS:2012aa,Chatrchyan:2012ta}. The sQGP was present in the first microsecond after the Big Bang, and also may be present during various events of astrophysical importance, such as supernovae and collisions of neutron stars. Hence, the exploration of this matter is of utmost importance with respect to our understanding of the smallest and largest scales of the Universe~\cite{Sorensen:2023zkk}.

Femtoscopy stands as a crucial tool in heavy-ion physics, enabling the exploration of the femtometer and yoctosecond space-time structure of the created medium (see, e.g., Refs.~\cite{Lisa:2005dd} and~\cite{Verde:2006dh} for reviews at high and intermediate energies, respectively). This is in particular useful when investigating the phase structure of this matter, and the transition from sQGP to hadron gas~\cite{Bzdak:2019pkr}. It was observed recently in various experiments from SPS~\cite{NA61SHINE:2023qzr,Porfy:2024ohy,Porfy:2024kbk} through RHIC~\cite{PHENIX:2017ino,Mukherjee:2023hrz,Kovacs:2023qax,Kincses:2024sin,PHENIX:2024vjp} to LHC~\cite{Korodi:2023fug,CMS:2023xyd} that contrary to expectation based on a small mean free path and the central limit theorem, the freeze-out two-particle spatial distribution does not follow a Gaussian, but is characterised by long tails and L\'evy-stable distributions. There are several phenomena in high-energy heavy-ion collisions that may be responsible for this~\cite{Csanad:2024hva}, but the most general reason might be the fact that the evolution of hadrons can be described by a L\'{e}vy walk. We investigate this hypothesis in this manuscript. The importance of exploring the shape of particle freeze-out distributions with femtoscopy lies in their connection to the dynamics and phase structure of the sQGP~\cite{Pratt:1984su,Lisa:2005dd,Wang:2024yke,Sorensen:2023zkk,Lacey:2014wqa} and understanding hadronic interactions~\cite{STAR:2015kha,ALICE:2020mfd,Tolos:2020aln,Fabbietti:2020bfg,Rzesa:2024oqp,Stefaniak:2024fkf}.

We note that the deviation from Gaussian distributions has been extensively studied in heavy-ion collisions at lower energies, where the heavy tail is mostly attributed to secondary decay emission of particles (resonance decays or evaporative emissions by hot systems). Such investigations have been reported on in Refs.~\cite{Brown:2000aj,Verde:2001md,Verde:2003cx,Nzabahimana:2023tab} at lower energies to experimentally reveal a non-Gaussian source. In these works, instead of using a-priori analytical functions to fit the correlations functon, the Koonin-Pratt equation was directly inverted using imaging techniques~\cite{Brown:2000aj,Verde:2001md,Verde:2003cx} or the more recent deblurring technique~\cite{Nzabahimana:2023tab}.

\section{Results}
\subsection{Lévy walk in hadronic scattering}\label{ss:urqmdresults}

Lévy processes~\cite{Metzler:2009springerbook,Nolan:Levy} are a form of random walk that appear in Nature for a wide range of phenomena, from foraging of animals~\cite{Sims:2008nature,Reynolds:2010JRSI,Edwards:2007nature,Humphries:2010nature} and swarm dynamics~\cite{Reynolds:2016scirep} through various microbiological~\cite{Harris:2012nature,Ariel:2015natcomm,Reijers:2019natcomm}, chemical~\cite{Sokolov:1997prl}, and physical processes~\cite{Barthelemy:2008nature} to climate change~\cite{Ditlevsen:1999geophysreslett}. These types of random walks occur when the individual random variables, whose sum constitutes the final investigated random variable, do not follow a distribution with a finite second moment. The divergence of the second moment happens due to the heavy tails of the realised distributions. In this case, the central limit theorem is no longer valid, but the generalised central limit theorem may still apply~\cite{Gnedenko:GCLT}, leading to the general class of stable distributions (also called Lévy-stable, or alpha-stable)~\cite{Nolan:Levy} for the sum of the individual random variables. A defining property of these types of distributions is that a linear combination of two independent random variables with this distribution also follows the same stable distribution, up to a rescaling~\cite{Nolan:Levy}. Moreover, one can make a distinction between Lévy flight and Lévy walk~\cite{Shlesinger:1986springerbook,Dybiec:2017PRD}: the latter is a special case when there is a finite speed for the individual steps in the random walk, hence longer steps take a larger time to happen. In many physical systems, the latter are applicable.

To investigate if these phenomena could also appear in heavy-ion collisions, we first turn to hadronic scattering in the Ultra-Relativistic Quantum Molecular Dynamics (UrQMD) framework to determine step length distributions (see details about the model in Methods). While the applicability of this model in ultra-relativistic heavy-ion collisions is limited, it is realistic for lower energies and provides a comprehensive picture of hadron scattering in the late stages of high-energy collisions. It models the collision by starting from the constituents of the colliding nuclei, letting them undergo an interaction, where one or more new particles might be created. These ``second-generation'' particles then also interact (or decay), and this chain of interactions proceeds until the system expands to a stage where collisions are not possible anymore, or in the practice of UrQMD until a large enough preset time. The four main types of interactions are:
\begin{itemize}
\item Scattering ($2\rightarrow 2$ process, i.e. a 2-by-2 scattering, elastic or inelastic)
\item Decay ($1\rightarrow N$ process with $N=2,3$, i.e., 2 or 3 particles are created from one)
\item Coalescence ($2\rightarrow 1$ process; also called 'annihilation' in UrQMD)
\item String creation and subsequent fragmentation ($2\rightarrow N$ process, with $N\gg 2$ usually)
\end{itemize}

\begin{figure}
    \centering
    \begin{tikzpicture}
        \node[anchor=south west,inner sep=0] (img) at (0,0) {\includegraphics[width=0.495\textwidth,trim={90 10 50 60},clip]{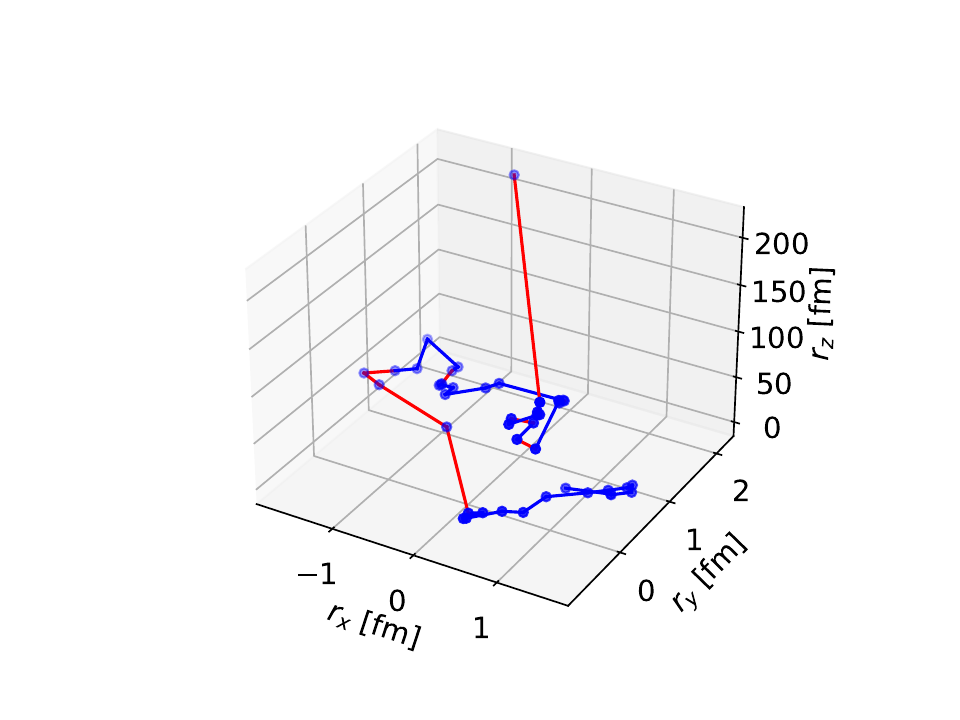}};
        \node[anchor=north west] at (0.1,0.95) {(a)};
    \end{tikzpicture}
    \vspace{1em} 
    \begin{tikzpicture}
        \node[anchor=south west,inner sep=0] (img) at (0,0) {\includegraphics[width=0.495\textwidth,trim={90 10 50 60},clip]{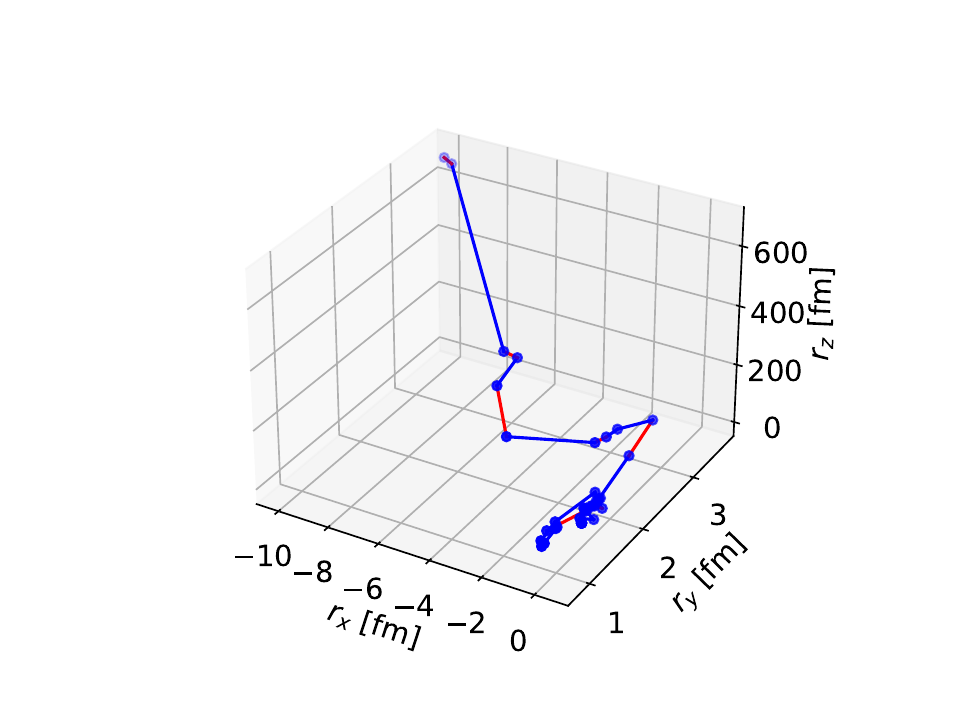}};
        \node[anchor=north west] at (0.1,0.95) {(b)};
    \end{tikzpicture}
    \begin{tikzpicture}
        \node[anchor=south west,inner sep=0] (img) at (0,0) {\includegraphics[width=0.495\textwidth,trim={90 10 50 60},clip]{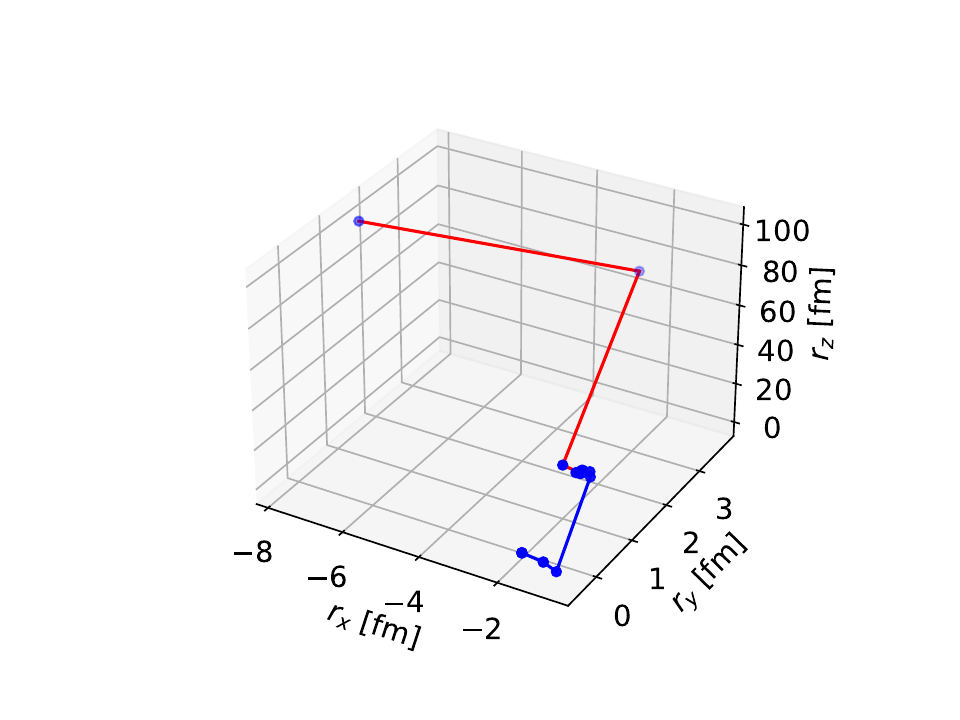}};
        \node[anchor=north west] at (0.1,0.95) {(c)};
    \end{tikzpicture}
    \caption{History of a three example pions in 200 GeV UrQMD Au+Au collisions. The blue dots show interactions, the red lines show pion propagation, and the blue lines show the propagation of other types of particles. The timelines in each panel (a, b, c) start from a nucleon in a collision and end in a final-state pion.}
    \label{f:urqmdpaths}
\end{figure}

Using the special output of UrQMD that provides the entire history of interactions, we analysed 100 gold-gold (Au+Au) collisions at $\sqrt{s_{\rm NN}}=200$ GeV center-of-mass per nucleon pair energy (corresponding to top RHIC energy for heavy-ions). The maximum impact parameter of the collisions was set 4.7 fm, corresponding to the $0-10\%$ centrality class. We selected final pions (the most abundant particles in such collisions) at their last point of scattering, also called kinetic freeze-out, as particle momenta cease to change (``freeze'') at this point. We then tracked these pions back to the constituents of the colliding nuclei. A few selected paths are shown in Fig.~\ref{f:urqmdpaths}. These illustrate that before a pion arrives at its last point of scattering and starts its flight to the detectors, it (or its predecessors) undergoes many steps. These paths resemble Lévy-flights or Lévy-walks as shown in, e.g., Refs.~\cite{Metzler:2009springerbook, Reynolds:2016scirep,Harris:2012nature,Ariel:2015natcomm,Reijers:2019natcomm}.

What we are interested in is the distribution of points where detected pions start their straight flight toward the detectors---this we call the source distribution. The random variable representing the location of the particle at freeze-out can then be understood as the vector sum of the individual steps. Due to the central limit theorem, if these steps follow a distribution that has a finite second moment, then the freeze-out coordinates follow a Gaussian distribution. On the other hand, if the length distribution of the steps does not have a finite second moment, then the central limit theorem cannot be applied. However, the generalised central limit theorem can still be utilised, and in this case, the freeze-out distribution will follow a power-law-tailed Lévy-stable distribution. To understand which of these cases apply to these collisions, we calculate the step length distributions separated into the four types of processes listed above. Here, the step length represents the distance covered by the particle \emph{before} the actual process. This is directly related to the mean free path of the given particle with respect to the given process, which in turn is given by the density of particles and the cross-section of the given process.

\begin{figure}
    \centering
    \includegraphics[width=0.9\textwidth]{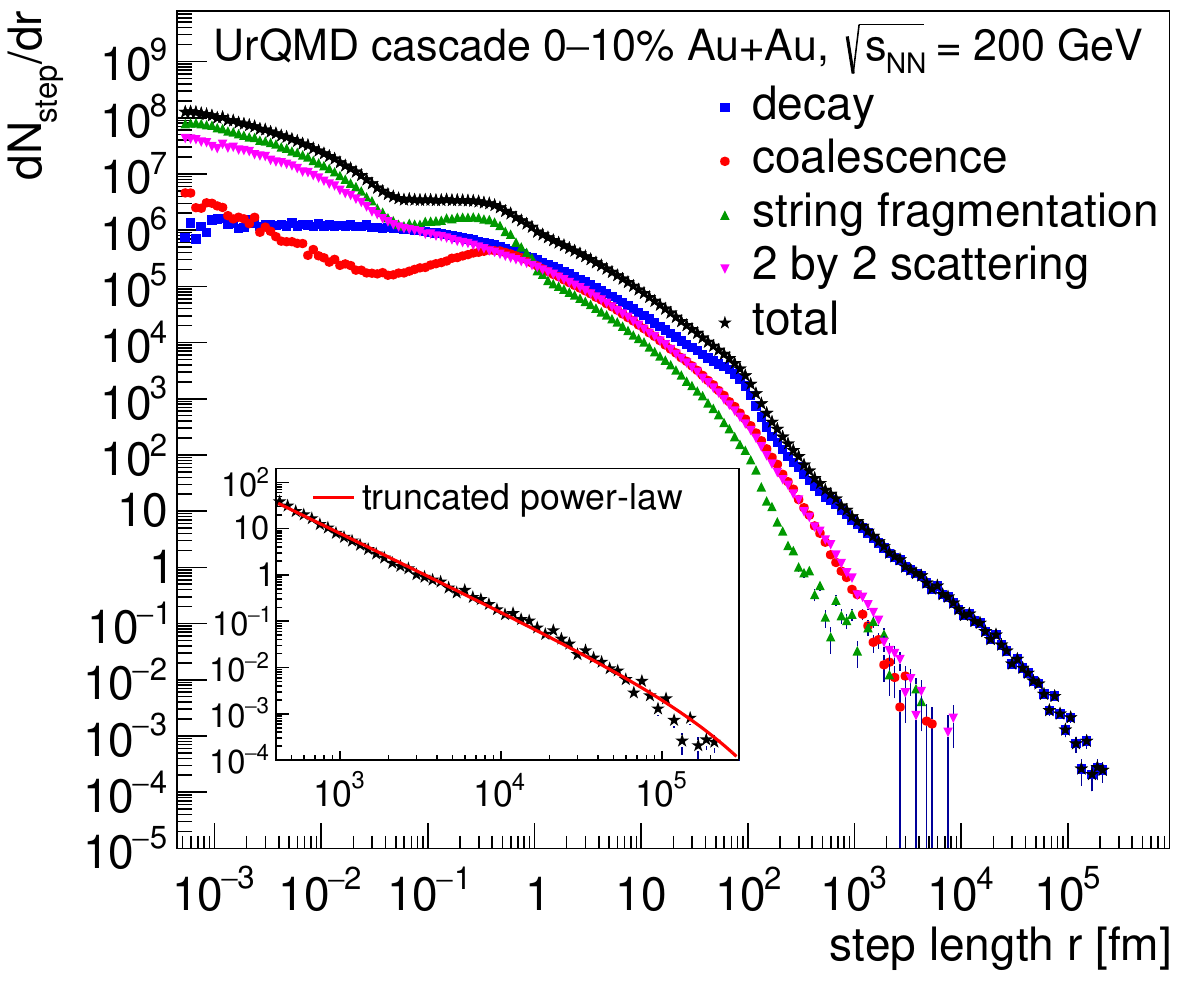}
    \caption{Step length distribution in 200 GeV UrQMD Au+Au collisions. Black stars mark the total distribution, and various colors indicate its components (decay: blue squares, coalescence: red circles, string fragmentation: green upward facing triangles, 2 by 2 scattering: magenta downward facing triangles). In a small inset, the tail of the total distribution is shown, fitted with a truncated power-law function.}
    \label{f:stepsizedist}
\end{figure}

The step length distributions are shown in Fig.~\ref{f:stepsizedist}. The total distribution appears to possess a power-law tail due to the decay processes of long-lived resonances producing pions. The regions below approximately 100 fm step length are not dominated by a single type of process, but all contribute to it. For the smallest step lengths (below 0.01 fm), string fragmentation and scattering are the most important. One furthermore observes a strong similarity between coalescence and scattering step length distributions above 1 fm. At the largest step length, decays dominate the total distribution, but all of the observed distributions exhibit a heavy tail. This may be due to the expansion of the medium: as time passes, the density decreases, and hence mean free paths increase. Thus, during the next step, even more time may pass (on average), ultimately leading to these power-law-tailed distributions. In particular, the tail of the total step-size distribution, shown in the inset of Fig.~\ref{f:stepsizedist}, exhibits a power-law exponent of $-1.69$.
In order to characterise the tail, following the methodology established in Ref.~\cite{Clauset:2009aaa}, we compared our results to other candidate distributions, in particular multi-exponential distributions, via the log-likelihood method, using the Akaike information criterion~\cite{Akaike:1974aaa}. Goodness-of-fit tests revealed that, indeed, the truncated power-law-tailed distribution gives the best description. These tests are detailed in the Methods section.
As suggested in Ref.~\cite{Stumpf:2012aaa}, power-law distributions should apply over at least two orders of magnitude along both axes to be convincing about their power-law-like nature. This criterion is fulfilled in our case.
A power-law distribution with the obtained exponent of $-1.69$ does not have a second moment---in reality, due to the finite time of the evolution, a cutoff in the form of an exponential truncation appears (as shown in the inset of Fig.~\ref{f:stepsizedist}).
We also note that the resulting single-particle freeze-out distribution, as expected, has a tail with the same power-law exponent.
It is furthermore important to point out that the above explanation for the heavy tails of pion distributions in ultrarelativistic collisions is in line with similar to studies at lower energies~\cite{Brown:2000aj,Verde:2001md,Verde:2003cx,Nzabahimana:2023tab}.

\begin{figure}
    \centering
    \includegraphics[width=0.6\textwidth]{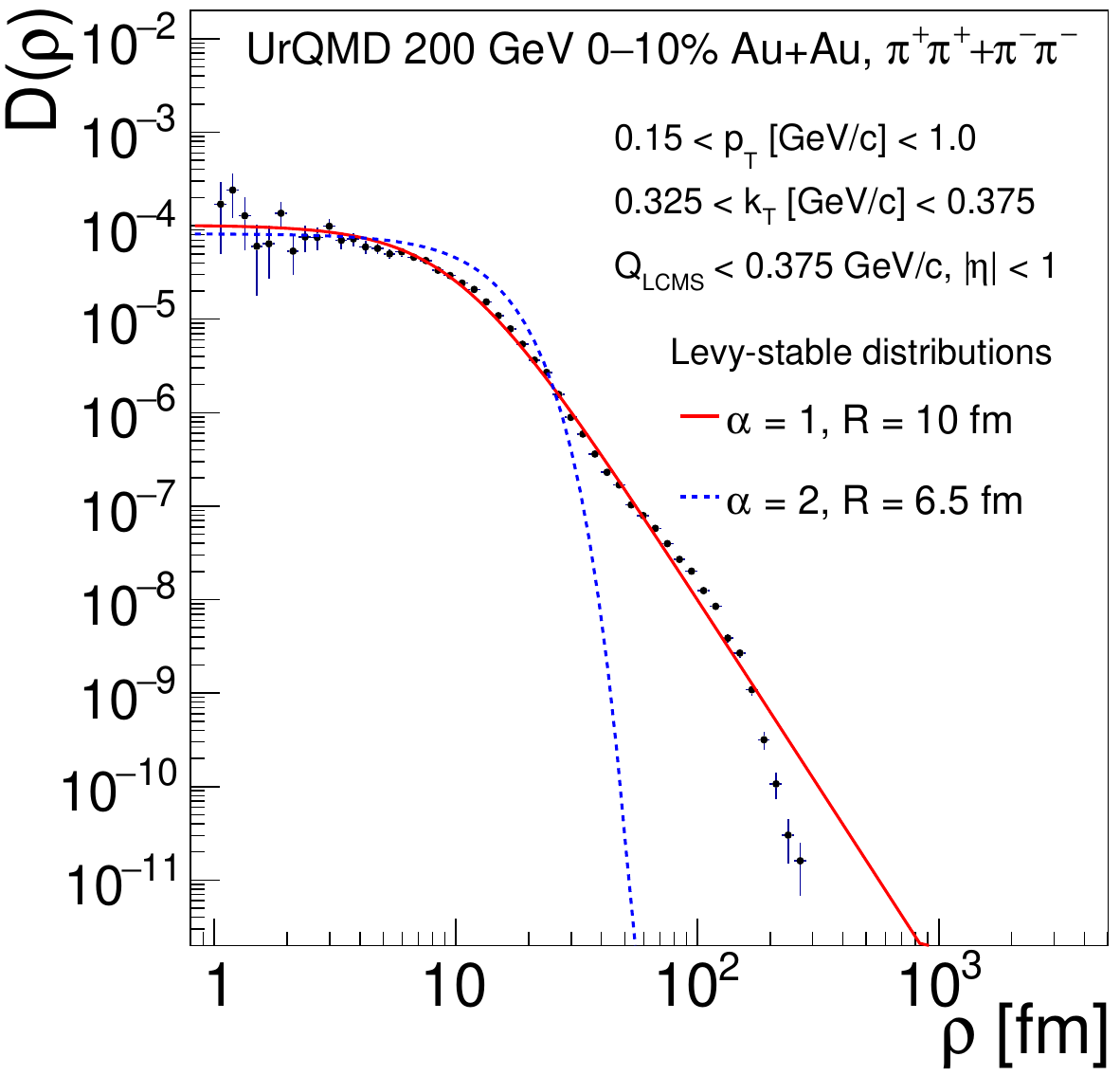}
    \caption{Identical pion distance distribution in a 200 GeV Au+Au event (points with statistical error bars) and two Lévy-stable distributions: one with $\alpha=1$ and $R=10$ fm (solid red line) and one with $\alpha=2$ and $R=6.5$ fm (dashed blue line) to understand the main features of the simulated result. The utilized kinematic cuts are also indicated on the plot. Error bars represent statistical uncertainties, stemming from the square root of the number of pairs in the given bin.}
    \label{f:urqmdDrho}
\end{figure}

These observations suggest that, indeed, the freeze-out distributions may be Lévy-stable. To this end, we investigate these freeze-out distributions for various cases. In particular, we calculate the angle-averaged freeze-out distance distribution $D(\rho)$, where $\rho$ is the distance variable. This distribution is equal to the autoconvolution of the individual freeze-out distributions (as defined in the Methods section) and is hence less prone to fluctuations within individual events (due to the smoothing nature of the autoconvolution operation, which averages out fluctuations by mixing values from different regions). More importantly, it allows one to investigate distributions event-by-event, as the number of pairs (each of which yields one entry into the $D(\rho)$ distribution) is proportional to multiplicity squared. This is especially relevant for pions, which are the most abundant particles in high-multiplicity central heavy-ion collisions. Such a $D(\rho)$ distribution for one individual example event is shown in Fig.~\ref{f:urqmdDrho}, with kinematic cuts close to experimental analyses (see, for example, Refs.~\cite{NA61SHINE:2023qzr,Porfy:2024ohy,Porfy:2024kbk,PHENIX:2017ino,Mukherjee:2023hrz,Kovacs:2023qax,Kincses:2024sin,PHENIX:2024vjp,Korodi:2023fug,CMS:2023xyd}). Since in experiment spatial distributions are not accessible, one utilizes momentum correlations fitted with a functional form calculated from the assumed source distribution to infer the spatial properties of the system (see the Methods section for details). To make our distribution comparable with that extracted from experiments, we apply experimental cuts on single particle momentum and pair momentum. As apparent in Fig.~\ref{f:urqmdDrho}, and as expected based on the distribution of individual step lengths, the freeze-out distribution is far from a Gaussian and close to a spherically symmetric Lévy-stable distribution, defined as
\begin{align}
\mathcal{L}(\alpha,R,\vec{r})=\frac{1}{(2\pi)^3}\int d^3\vec{q}\, e^{i\vec{q}\cdot\vec{r}} e^{-\frac{1}{2}|\vec{q}R|^{\alpha}},\label{eq:Levydef}
\end{align}
where $\vec{q}$ is the integration variable, $\vec{r}$ is the variable of the distribution, $R$ is called the Lévy-scale, characterizing the size, and $0<\alpha\leq 2$ is called the Lévy-index characterizing the shape. We omit the $\alpha > 2$ parameter range: although the function is defined here as well, it is not always positive. In the special case of $\alpha=2$, the distribution is a Gaussian, while in case of $\alpha < 2$ it exhibits a power-law tail with exponent $-3{-}\alpha$.

As Fig.~\ref{f:urqmdDrho} shows, the observed $D(\rho)$ from UrQMD is qualitatively compatible with a Lévy-stable distribution with $\alpha=1$. This provides another argument showing that the final-state positions of pions in a hadron gas simulation are sums of steps converging to L\'{e}vy-stable distributions. In order to compare to experimental observations in high-energy heavy-ion collisions, we have to investigate a model incorporating all stages and aspects of these collisions.

\subsection{Lévy-stable distributions in a complete model calculation}\label{ss:eposresults}

As a second step, we investigate the EPOS (``Energy conserving quantum mechanical multiple scattering approach, based on Partons, Off-shell remnants, and  Splitting of parton ladders'') model framework, a complete, state-of-the-art simulation of ultra-relativistic heavy-ion collisions. Compared to UrQMD, the main difference in such a more realistic model is that hadrons are created only after a hydrodynamically evolving sQGP phase. Hence, hadronic scattering starts from a thermodynamically equilibrated state and lasts only for a shorter time, allowing for less diffusion and fewer steps in a Lévy-walk. Thus, a Lévy-index $\alpha$ closer to the Gaussian $\alpha=2$ is expected in such a case.

\begin{figure}
    \centering
    \begin{tikzpicture}
        \node[anchor=south west,inner sep=0] (img) at (0,0) {\includegraphics[width=\textwidth]{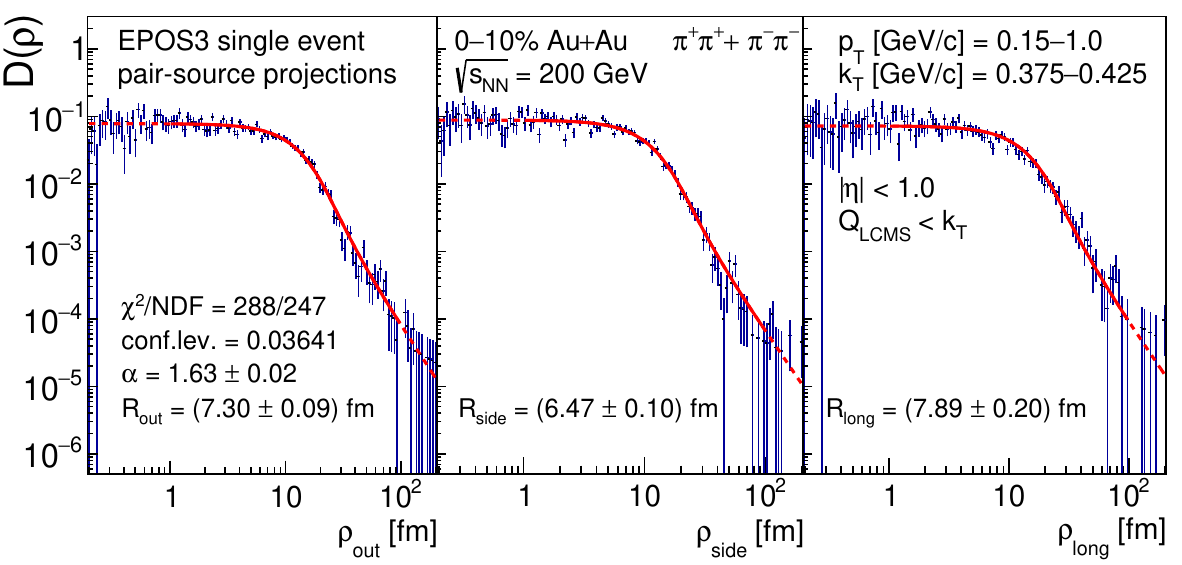}};
        \node[anchor=north west] at (4.,4.) {(a)};
        \node[anchor=north west] at (8.1,4.) {(b)};
        \node[anchor=north west] at (12.2,4.) {(c)};
    \end{tikzpicture}
    \caption{Example pair freeze-out distance distributions from one EPOS event, fitted by projections of Lévy-distributions. The utilized kinematic cuts are denoted on panel (c). The one-dimensional projections of the measured pair-source distribution for the out direction (a), side direction (b) and long direction (c) are denoted by blue markers with error bars. The red lines denote the corresponding one-dimensional projections of the three-dimensional elliptically contoured Lévy-stable distribution. Error bars represent statistical uncertainties, stemming from the square root of the number of pairs in the given bin.}
    \label{f:EPOSexamplefit}
\end{figure}

\begin{figure}
    \centering
    \begin{tikzpicture}
        \node[anchor=south west,inner sep=0] (img) at (0,0) {\includegraphics[width=0.495\textwidth]{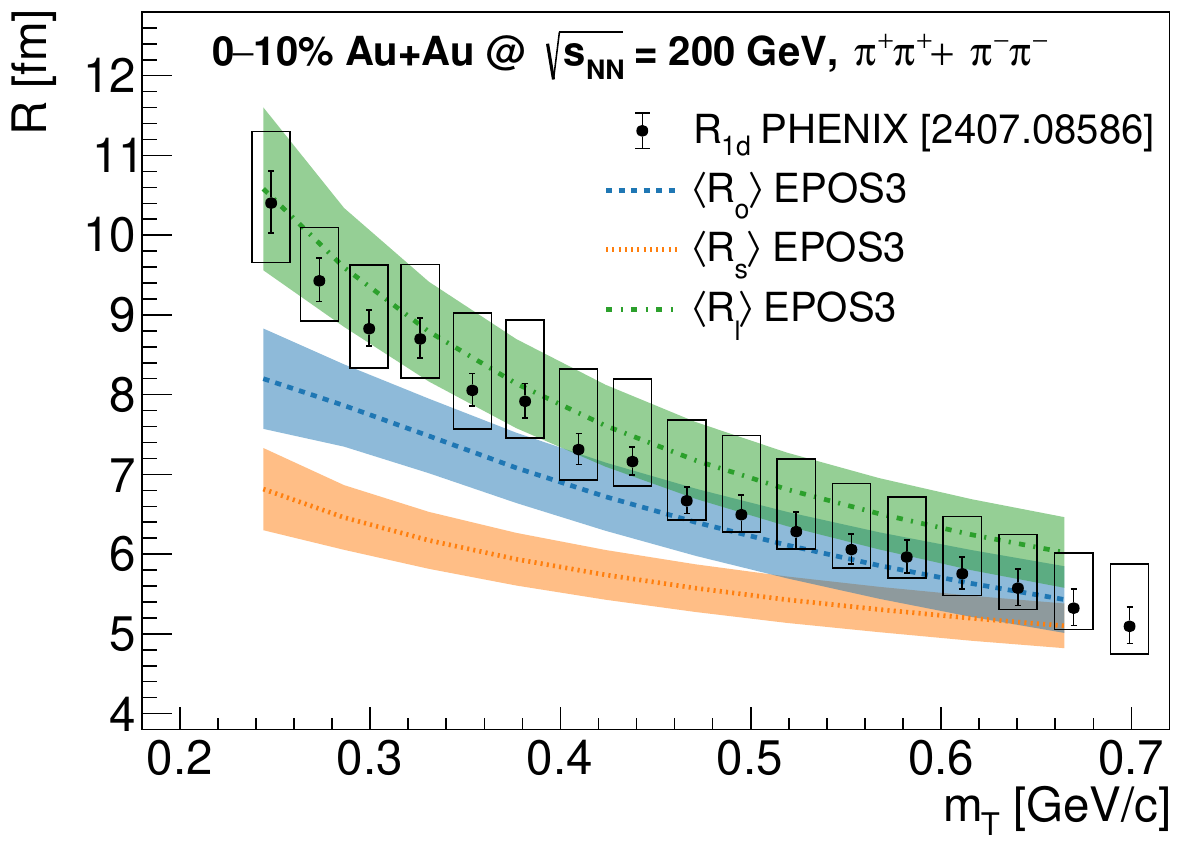}};
        \node[anchor=north west] at (0.9,1.3) {(a)};
    \end{tikzpicture}
    \vspace{1em} 
    \begin{tikzpicture}
        \node[anchor=south west,inner sep=0] (img) at (0,0) {\includegraphics[width=0.495\textwidth]{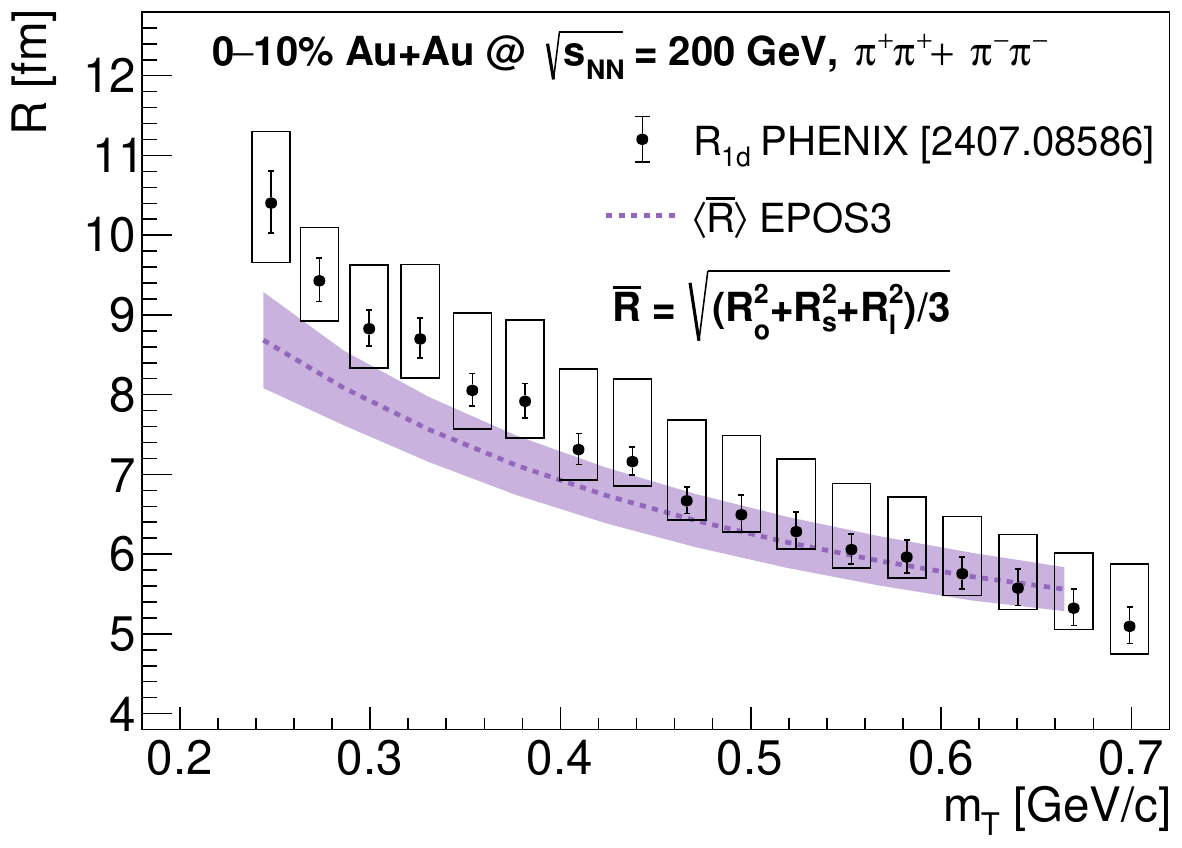}};
        \node[anchor=north west] at (0.9,1.3) {(b)};
    \end{tikzpicture}
    \begin{tikzpicture}
        \node[anchor=south west,inner sep=0] (img) at (0,0) {\includegraphics[width=0.495\textwidth]{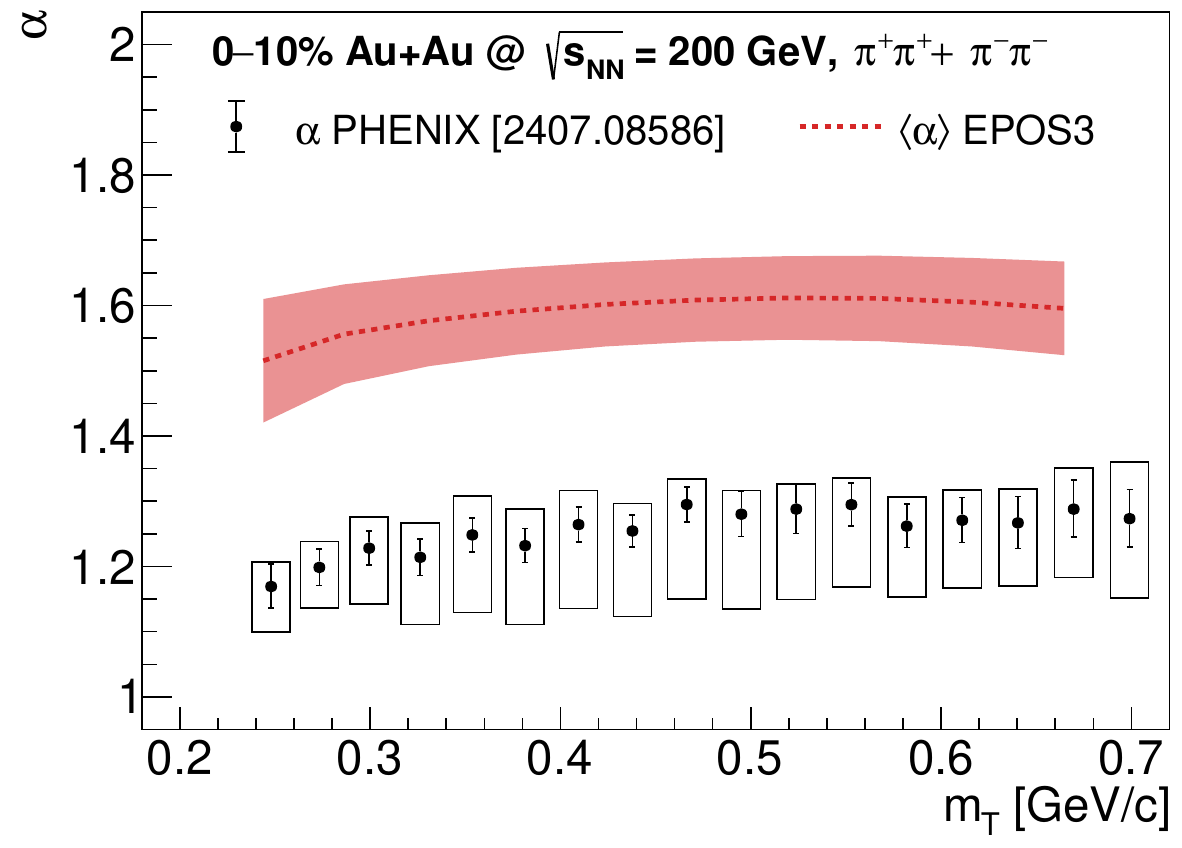}};
        \node[anchor=north west] at (0.8,1.3) {(c)};
    \end{tikzpicture}
    \caption{Lévy source parameters versus pair transverse mass $m_{\rm T}$. The panels show the  experimental values from the PHENIX Collaboration~\cite{PHENIX:2024vjp}, compared to the directional EPOS radii (a), an averaged radius (b), as well as the exponent $\alpha$ from experiment and simulation (c). For the simulated results, the dashed lines denote the mean and the colored regions denote the variance of the given parameter across all event-by-event fits. For the experimental data, error bars denote statistical uncertainties, while boxes represent systematical uncertainties.}
    \label{f:EPOSkTdep}
\end{figure}

We simulated $\sqrt{s_{\rm NN}}=200$ GeV Au+Au collisions with EPOS (see details in Methods), all belonging to the class of 10\% most central collisions (in terms of impact parameter of the colliding nuclei). Based on this sample, we investigated event-by-event distance distributions of pairs and fitted them with elliptically contoured, three-dimensional Lévy distributions defined as~\cite{Nolan:2013compstat}
\begin{align}
\mathcal{L}(\alpha,R^2,\vec{r})=\frac{1}{(2\pi)^3}\int d^3\vec{q}\, e^{i\vec{q}\cdot\vec{r}} e^{-\frac{1}{2}|\vec{q}^T R^2 \vec{q} |^{\alpha/2}},\label{eq:Levy3Ddef}
\end{align}
where again $\alpha$ is the Lévy-index, $\vec{r}$ is the variable of the distribution, $\vec{q}$ is an integration variable, and now $R^2={\rm diag}(R^2_{\rm out},R^2_{\rm side},R^2_{\rm long})$ is the matrix containing the Lévy-scale parameters, in the Bertsch-Pratt coordinate system (detailed in the  Methods section). Fig.~\ref{f:EPOSexamplefit} shows an example fit for one event with the projections of both simulated and fitted distributions. As customary both in experimental analyses and in model calculations, we assume that the parameters of the pair distance distribution $D(\vec \rho)$ may depend on the kinematic characteristics of the pairs, in particular on the average transverse momentum $k_{\rm T}$ (i.e. the projection of the average momentum perpendicular to the direction of the colliding nuclei). In this spirit, given the event-by-event fits to the distance distributions, we extract the source parameters $\alpha$, $R_{\rm out}$, $R_{\rm side}$, and $R_{\rm long}$, as a function of the average transverse mass of the pair (defined as $m_{\rm T} = \sqrt{k_{\rm T}^2+m_\pi^2}$, where $m_\pi$ is the pion mass), as shown in Fig.~\ref{f:EPOSkTdep}. These are then compared to experimental data from the PHENIX (Pioneering High Energy Nuclear Interaction eXperiment) Collaboration~\cite{PHENIX:2024vjp}, in the same collision system (10\% most central, $\sqrt{s_{\rm NN}}=200$ GeV Au+Au collisions; a thorough comparison for all measured centralities is the subject of a follow-up paper). The Lévy index is significantly different in simulations as compared to the data, but both are far from the Gaussian case. The scale parameters from the simulations are closer to the experimentally observed ones. Note that only a spherically symmetric radius has been extracted by PHENIX, thus besides the three directional radii, we also compare an average radius, ${\Bar{R}=\sqrt{\smash[b]{\big( R_{\rm out}^2+R_{\rm side}^2+R_{\rm long}^2\big)/3}}}$, which is quite close to the measurements. These observations show that three-dimensional Lévy-stable distributions appear in EPOS, and while they are close in size, they are a bit further in shape from what is observed in the experiment. This indicates that besides Lévy walks, other phenomena also may play a role in a real heavy-ion collision, or the effect of Lévy walk is stronger in reality as in EPOS, closer to the situation in a pure UrQMD setting.

\section{Discussion}\label{s:discussion}

We have shown that Lévy walk appears in hadronic rescattering and decays after a high-energy heavy-ion collision, and leads to power-law tailed spatial freeze-out pair distributions. In order to understand the dynamics of the sQGP and hadron interactions, it is of utmost importance to explore the shape of these distributions. Our observations
indicate that, indeed, Lévy-processes occur in high-energy heavy-ion collisions. Our study points towards more appropriate ways of extracting information on spatial particle distributions from data, based on Lévy-stable distributions. In this methodology, as detailed in Refs.~\cite{NA61SHINE:2023qzr,Porfy:2024ohy,Porfy:2024kbk,PHENIX:2017ino,Mukherjee:2023hrz,Kovacs:2023qax,Kincses:2024sin,PHENIX:2024vjp,Korodi:2023fug,CMS:2023xyd}, one assumes a Lévy-stable distributions for the particle-emitting source (while  also allowing for and investigating possible deviations, for example via the Lévy-expansions~\cite{Novak:2016cyc}), and extracts the parameters of this distribution.

To compare our ideas to experimental results, we investigated a detailed, state-of-the-art simulation of these collisions and found that on an event-by-event basis, the freeze-out pair distributions can be described by a three-dimensional Lévy-stable source. As of now, no three-dimensional experimental analysis utilising Lévy-type sources has been published, only angle-averaged measurements. Hence, this three-dimensional model study is the first to show that the reason behind the appearance of such source shapes is not the event- nor the angle-averaging. The extracted source parameters, Lévy index $\alpha$, and Lévy scales $R_{\rm out,side,long}$ were also compared to experimental results. The Lévy scales are compatible with a recent measurement from the PHENIX Collaboration~\cite{PHENIX:2024vjp}, however, the Lévy index is significantly larger, albeit still far from the Gaussian case. The difference may be attributed to the lack of long-range Coulomb interaction in the model---this has been shown to lead to Lévy flight~\cite{Kiselev:2021lar}. Altogether, our results may signal a strong effect of Lévy walks in yet another realm of Nature.

\section{Methods}\label{s:methods}

\subsection{Momentum correlations and spatial distributions}\label{ss:DrCq}

In heavy-ion collision experiments, the femtometer-scale locations and distances are not directly accessible. Instead, one utilizes the momentum correlation function, defined as
\begin{align}\label{e:c2def}
C_2(p_1,p_2)=\frac{N_2(p_1,p_2)}{N_1(p_1)N_1(p_2)},
\end{align}
where $p_1$ and $p_2$ are the four-momenta of the detected particles, $N_2(p_1,p_2)$ is the two-particle momentum distribution while $N_1(p_1)$ and $N_1(p_2)$ are the single-particle momentum distributions. The division by these removes kinematic and phase-space effects, and the resulting $C_2(p_1,p_2)$ represents the true correlations. There are usually multiple reasons for correlated particle production: collective flow, jets, resonance decays, momentum conservation; but in high-multiplicity heavy-ion collisions, the main source of correlations at low relative momentum is usually femtoscopic in nature. This includes the quantum-statistical (for pions, Bose-Einstein) correlations, as well as ones stemming from final state interactions. In a Wigner-function formalism, assuming chaotic (thermal) particle production, the two-particle momentum distribution can be calculated~\cite{Makhlin:1987gm,Pratt:1990zq,Csorgo:1999sj,PHENIX:2017ino} according to the Yano-Koonin formula~\cite{Yano:1978gk} as
\begin{align}\label{e:yanokoonin}
N_2(p_1,p_2) = \int d^4x_1 d^4x_2 S(x_1,p_1) S(x_2,p_2) |\psi_{p_1,p_2}(x_1-x_2)|^2,
\end{align}
where $|\psi_{p_1,p_2}(x_1-x_2)|^2$ is the squared modulus of the pair wave-function, which is symmetric for identical bosons and antisymmetric for identical fermions, and depends also usually on the final-state interaction between the members of the pair. Furthermore, $S(x,p)$ is the probability density of creating a particle with four-momentum $p$ at space-time point $x$---this is usually called the freeze-out distribution.

Utilizing the so-called smoothness approximation~\cite{Lisa:2005dd,Csorgo:1999sj,Wiedemann:1999qn}, i.e. assuming that the momentum difference is much smaller than the mean momentum, Eq.~\eqref{e:c2def} can be rewritten as
\begin{align} 
C_2(q,K) = \int dr D(r,K) |\psi_{q}(r)|^2, \label{e:C2D}
\end{align}
where we now define the average pair momentum $K=\frac12(p_1+p_2)$ and relative momentum $q=p_1-p_2$. In case of plane-waves the wave-function term becomes $\exp(iqr)$, with $qr$ representing the Lorentz-product of $r$ and $q$. In other words, this approximation asserts that $p_1\approx p_1$ in the variables of the source function $S$ in Eq.~\eqref{e:yanokoonin}. In the above equation, we introduced the spatial pair distribution,
\begin{align}\label{e:S2Dexpr}
D(r,K) = \int S\left(x+\frac{r}{2},K\right) S\left(x-\frac{r}{2},K\right) d^4x,
\end{align}
where $r=x_1-x_2$ is the distance between particles in a pair, and $K$ is the (in the utilized approximation equal) momentum of each of the particles. Note that $D$ is the autoconvolution of $S$ in its first variable.

Eq.~\eqref{e:C2D} is the central equation of femtoscopy, indicating that by measuring the correlation function $C_2$ in momentum space, one can access information about the spatial distance distribution $D$ and the pair wave function $\psi$. Note that in this picture, the ``main'' variable of $C_2$ is $q$, and of $D$ it is $r$, and the dependence on mean momentum $K$ is assumed to be weaker and usually appears through the $K$-dependence of the parameters of $D$ or $C_2$. Note furthermore that at small relative momenta, the experimental correlation functions may be affected by distortions induced by detector efficiency, as seen in Fig. 3b of Ref.~\cite{NA61SHINE:2023qzr} and Fig. 3 of Ref.~\cite{CMS:2023xyd}. In experimental analyses, the affected momentum region is usually omitted from the fits~\cite{PHENIX:2017ino,NA61SHINE:2023qzr,CMS:2023xyd}, and thus the extracted experimental source parameters reflect the source well.

It is furthermore important to note that in the case of identical particle femtoscopy, it is usually possible to calculate $\psi$ exactly for the Coulomb interaction, and to a good approximation also including the strong interaction~\cite{Kincses:2019rug}. Thus the observed momentum correlations make it possible to infer the femtometer scale geometry of particle production. On the other hand, for unlike-particle femtoscopy (but also for identical particles with less known interactions), the investigated quantity is usually $\psi$, and through it, one learns about the interaction and its potential~\cite{STAR:2015kha,ALICE:2020mfd,Tolos:2020aln,Fabbietti:2020bfg}.

\subsection{Proper spatial variable}
\label{ss:variable}
In the above calculations, space-time Lorentz vectors $x$ and $r$ appear, and also four-momenta $p$, $q$ and $K$. We can separate the latter into time-like and spatial components (in a given frame) as $q=(q_0,\vec{q})$ and $K=(K_0,\vec{K})$. We then observe that for identical on-shell particles $p_1^2=p_2^2=m^2$, where $m$ is the mass of the given particle type. Thus, the following relation can be established:
\begin{align}
q_0 = \vec{q}\cdot\vec{\beta},
\end{align}
where $\vec{\beta} = \vec{K}/K_0$. This in turn means that $q_0$ is not an independent variable, i.e., it can be expressed in terms of the spatial components of $q$. Thus the Lorentz-product appearing in the two-particle wave-function (containing $r=(t,\vec{r})$) can be expressed as $qr = \vec q \cdot \vec \rho$~\cite{Cimerman:2017lmm,Lokos:2016fze}, if
\begin{align}
\vec{\rho} \equiv \vec{r} - \vec{\beta} t,
\end{align}
and thus $\vec \rho$ takes the place of the proper spatial variable in \eqref{e:C2D}, i.e., $D(\vec \rho)$ appears there instead of $D(r)$ (suppressing the dependence on $K$). Thus we investigate the dependence on the above three-vector $\vec{\rho}$ in three-dimensional calculations and its magnitude in one-dimensional calculations (and wherever we write $\rho$ as a scalar, the magnitude of $\vec{r} - \vec{\beta} t$ is understood). Furthermore, the angle-averaged distribution is defined as $D(\rho)=\frac{1}{4\pi}\int D(\vec \rho) \rho^2 d\Omega$, where $d\Omega$ is the solid-angle element, denoting the angular integration.

\subsection{Frame choice}\label{ss:frame}
Let us also discuss the choice of reference frame for the analysis. Besides the laboratory frame, a straightforward choice is the pair center-of-mass frame or Pair Co-Moving System (PCMS), defined as the frame where the total three-momentum of the pair vanishes. This is a pair coordinate system, defined separately for each pair. In this system, $\vec{K}=0$, thus for identical particles $q_0=0$. Hence $|\vec{q}|=q_{\rm inv}\equiv\sqrt{-q\cdot q}$, i.e., here the magnitude of the three-momentum difference equals the (Lorentz invariant) Lorentz length of the four-momentum difference. However, the particle production in relativistic heavy-ion collisions is highly non-spherical in this frame~\cite{PHENIX:2004yan,PHENIX:2009ilf,STAR:2004qya,PHENIX:2007grx}. As a remedy, one may utilise the Longitudinally Co-Moving System~\cite{Pratt:1990zq, PHENIX:2017ino}, where $K_{\rm long}=0$, i.e., the particle momenta in the longitudinal direction (that of the beam) are of equal magnitude and of opposite sign. In this frame, the correlation function is much more spherically symmetric~\cite{PHENIX:2004yan}. In addition, one often introduces the Bertsch-Pratt coordinates~\cite{Bertsch:1988db,Pratt:1986ev}, where the ``out'' direction is defined as that of the average momentum of the pair perpendicular to the longitudinal ``long'' direction (i.e., by the direction of the three-vector $(K_x,K_y,0)$), while the ``side'' direction is perpendicular to both the longitudinal and the out directions. In this frame and with these coordinates,
\begin{align}
K_{\rm long}&=K_{\rm side}=0\textnormal{, hence}\\
\vec{K}&=(K_{\rm out},0,0)\textnormal{, and }\\
\vec{\beta}&=(K_{\rm out}/K_0,0,0)\textnormal{, thus }\\
q_0 &= q_{\rm out}\beta_{\rm out}.
\end{align}
We utilize this frame and these coordinates throughout the manuscript, similarly to experimental analyses of Refs.~\cite{PHENIX:2017ino,CMS:2023xyd}.

In the Bertsch-Pratt coordinate system and the LCMS, the components of the above-mentioned spatial three-vector variable $\rho$ can be calculated with the lab-frame coordinates as
\begin{align}
    \rho_{\rm out}^{\rm LCMS} &= r_x\cos\varphi+r_y\sin\varphi-\frac{k_{\rm T}}{K_0^2-K_z^2}(K_0t-K_zr_z),\\
    \rho_{\rm side}^{\rm LCMS} &= -r_x\sin\varphi+r_y\cos\varphi,\\
    \rho_{\rm long}^{\rm LCMS} &= \frac{K_0r_z-K_zt}{\sqrt{K_0^2-K_z^2}},
\end{align}
where ${\cos\varphi = K_x/k_{\rm T}}$, ${\sin\varphi = K_y/k_{\rm T}}$, and ${k_{\rm T} = \sqrt{K_x^2+K_y^2}}$.

\subsection{Simulations}
In this subsection we indicate details of the simulations.

\subsubsection{UrQMD}\label{sss:urqmd}
For the Lévy walk study, we utilized the UrQMD transport model~\cite{Bass:1998ca, Bleicher:1999xi} (version 3.4). This model follows microscopic transport based on elastic and inelastic hadron scattering. It yields a fair description of various observables at low collision energies (well below the investigated top RHIC energy), and it provides an accurate picture of the hadron gas state and processes taking place after the hydrodynamic phase at high energies. For our study, we simulated 100 UrQMD Au+Au events at ${\sqrt{s_{\rm NN}}=200}$ GeV in the $0-10\%$ centrality class and investigated the individual ``history'' of final-state pions. We calculated individual step lengths and created their distributions on the basis of this history. Utilizing the freeze-out coordinate corresponding to the last point of interaction, we calculated freeze-out distance distributions, as discussed in the Results section.

\begin{figure}
    \centering
    \begin{tikzpicture}
        \node[anchor=south west,inner sep=0] (img) at (0,0) {\includegraphics[width=0.32\textwidth]{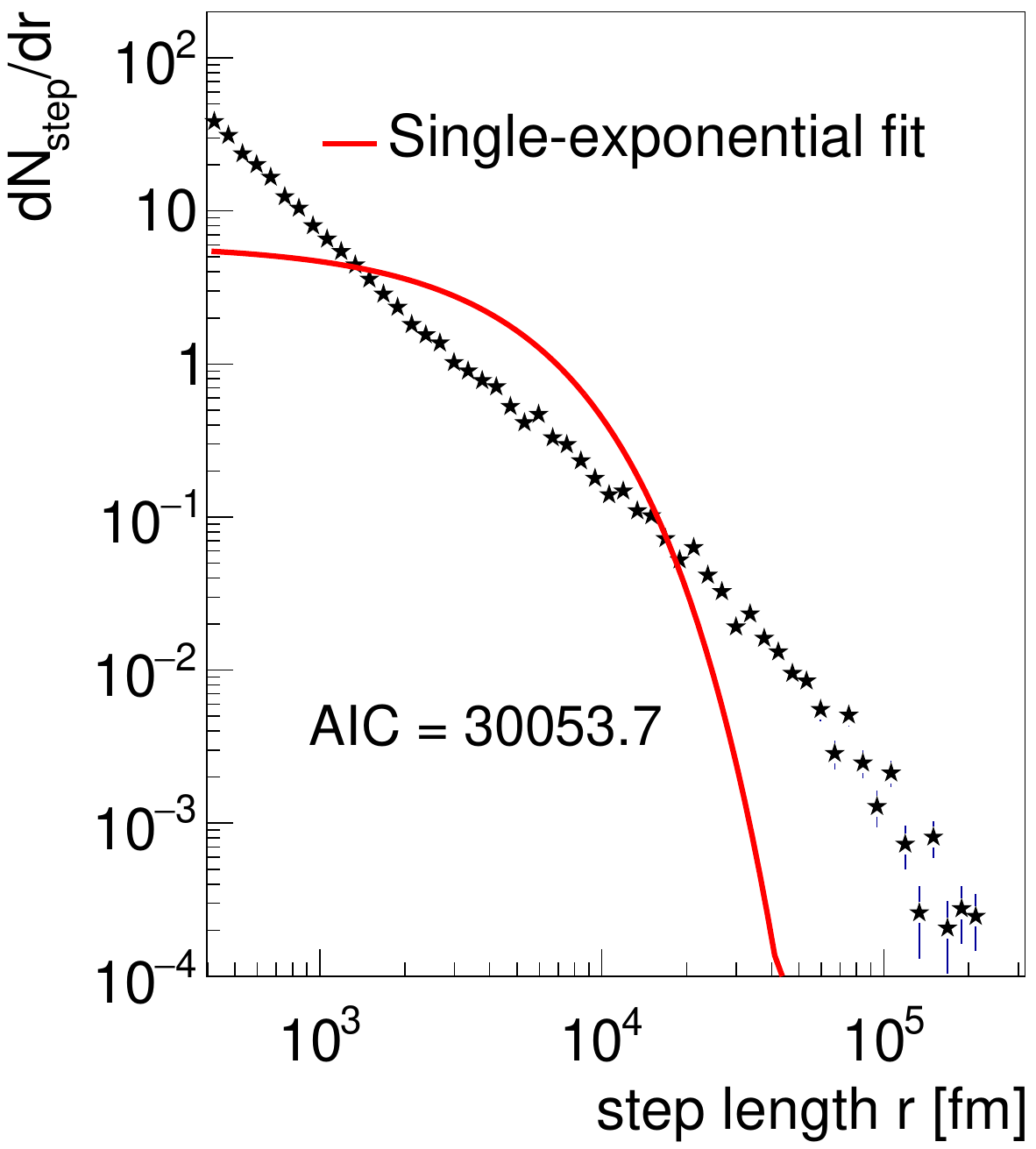}};
        \node[anchor=north west] at (0.9,1.4) {(a)};
    \end{tikzpicture}
    \begin{tikzpicture}
        \node[anchor=south west,inner sep=0] (img) at (0,0) {\includegraphics[width=0.32\textwidth]{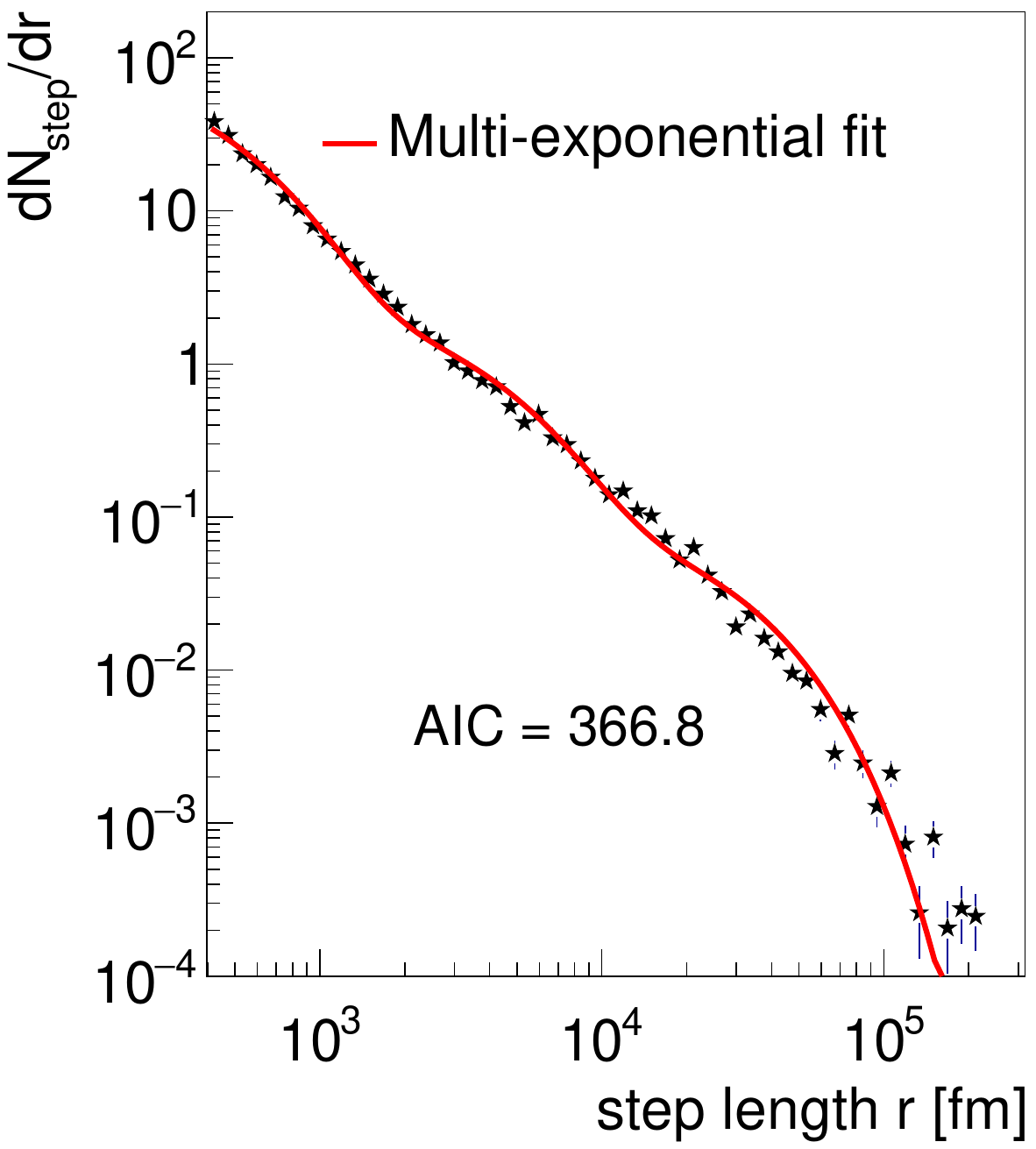}};
        \node[anchor=north west] at (0.9,1.4) {(b)};
    \end{tikzpicture}
    \begin{tikzpicture}
        \node[anchor=south west,inner sep=0] (img) at (0,0) {\includegraphics[width=0.32\textwidth]{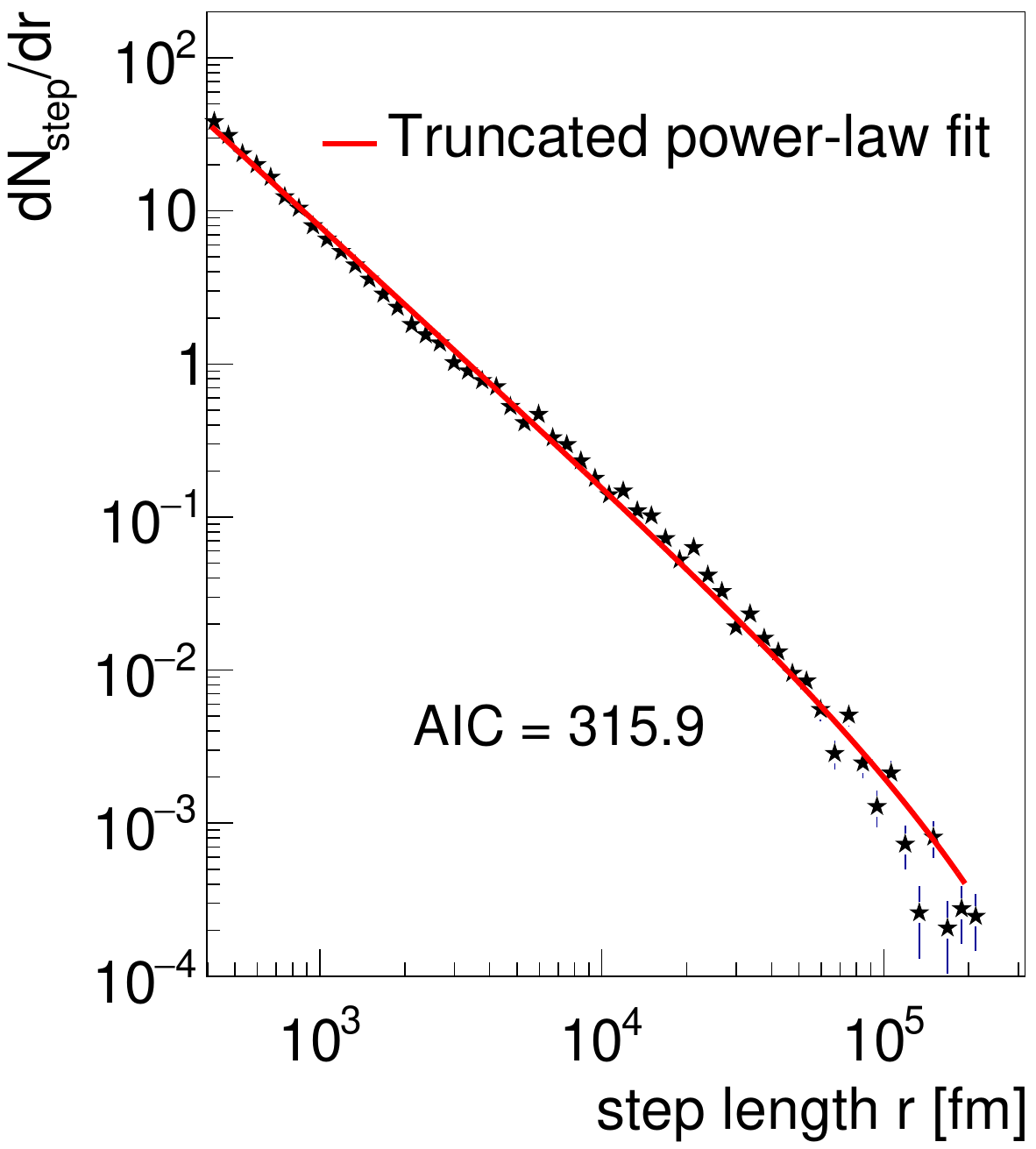}};
        \node[anchor=north west] at (0.9,1.4) {(c)};
    \end{tikzpicture}
    \caption{Step-length distribution versus candidate distributions. The three panels show the simulated step-length distribution (shown with black markers) fitted with a single exponential (a), a multi-exponential (b), and a truncated power law (c), in the range of 400 fm - 200000 fm, spanning more than two orders of magnitude. Based on the Akaike information criterion, the truncated power-law fit corresponds to the lowest AIC value, thus it provides the best description. Error bars represent statistical uncertainties, stemming from the square root of the number of pairs in the given bin.}
    \label{fig:akaiketest}
\end{figure}

In the case of the total step length distribution, we compared different candidate distributions via the log-likelihood method, using the Akaike information criterion~\cite{Akaike:1974aaa}, as shown in Figure~\ref{fig:akaiketest}. The fits covered the range of 400 fm to 200,000 fm, spanning more than two orders of magnitude. As apparent from the figure, a truncated power law provides a much better description than an exponential, and only a three-term multi-exponential gets close to its descriptive quality. Let us note that the ultimate AIC value, of course, depends on the statistics of simulated events. We now averaged 100 events; taking only a part of this sample reduces the log-likelihood term (increasing the actual likelihood). Thus, the absolute magnitude of the AIC values is not of central importance in the case of simulated data.

\subsubsection{EPOS}\label{sss:epos}
For the detailed analysis of high-energy heavy-ion collisions, we utilized the EPOS phenomenological model~\cite{Werner:2010aa} (version 359). This model reproduces all basic experimentally measured quantities in Au+Au collisions at the investigated RHIC energy of ${\sqrt{s_{\rm NN}}=200}$ GeV. In EPOS, a merged version of the Gribov-Regge theory and the eikonalized parton model is implemented~\cite{Drescher:2000ha}. In order to account for the variable local string density in different regions of the medium, string segments (based on local density as well as transverse momenta of the segments) are divided into the core and the corona~\cite{Werner:2010aa, Werner:2007bf, Werner:2013tya}. Subsequently, a 3+1 dimensional viscous hydrodynamic-based algorithm is applied~\cite{Werner:2010aa}; while the equation of state X3F was utilised, retaining compatibility with lattice calculations~\cite{Borsanyi:2010cj} at $\mu_{\text{B}}=0$. At the final stage, the Cooper-Frye formula~\cite{Cooper:1974mv} is utilized, and a hadronic afterburner is used, based on UrQMD~\cite{Bass:1998ca, Bleicher:1999xi}. For this paper, we investigated 30000 EPOS Au+Au events at $\sqrt{s_{\rm NN}}=200$ GeV, with an impact parameter vector uniformly distributed on a circle with a radius of 4.7 fm, corresponding to the 10\% most central collisions~\cite{Miller:2007ri}. We then investigated the freeze-out coordinates of particles, after the built-in UrQMD-based hadronic scattering and decays. We calculated the freeze-out pair distribution (distance distribution) $D(\vec{\rho})$, and fitted its projections with the corresponding Lévy functional forms, as discussed in the Results section.

\backmatter

\bmhead{Acknowledgements}

This research was funded by the NKFIH grants TKP2021-NKTA-64, PD-146589, K-146913, and K-138136. D. K. was also supported by the EKÖP-24 University Excellence Scholarship program of the Ministry for Culture and Innovation from the source of the national research, development, and innovation fund.

\bibliography{sn-bibliography}

\end{document}